\documentclass[letterpaper, 10 pt, conference]{IEEEtran}
\usepackage{url,amsmath,amsfonts,amssymb,bm, mathtools, algorithm,algpseudocode,mathrsfs,setspace,subfig,bbm,tikz,forest,pgfplots}

\IEEEoverridecommandlockouts  

\newtheorem{definition}{Definition}
\newtheorem{remark}{Remark}

\newtheorem{theorem}{Theorem}

\newcommand{\N}{{\mathbb N}}
\newcommand{\E}{{\mathbb E}}

\newcommand{\Py}{{\mathbb P}}

\newcommand{\vct}[1]{\bm{#1}}

\newcommand{\rv}[1]{\bm{\mathrm{#1}}}

\begin{document}

%\onecolumn
%\doublespacing
\title{Policy Design for Active Sequential Hypothesis
Testing using Deep Learning}

\author{Dhruva Kartik, Ekraam Sabir, Urbashi Mitra and Prem Natarajan
\thanks{
D. Kartik, and U. Mitra are with the Department of Electrical
Engineering, University of Southern California, Los Angeles, CA 90089
(e-mail: mokhasun@usc.edu; ubli@usc.edu).}
\thanks{
E. Sabir, and P. Natarajan are with the USC Information Sciences Institute, Marina Del Rey, CA 90292
(e-mail: esabir@isi.edu; pnataraj@isi.edu).}
% \thanks{This research was supported by }% <-this % stops a space
}

\maketitle

\thispagestyle{empty}
\pagestyle{empty}

\begin{abstract}
Information theory has been very successful in obtaining performance limits for various problems such as communication, compression and hypothesis testing. Likewise, stochastic control theory provides a characterization of optimal policies for Partially Observable Markov Decision Processes (POMDPs) using dynamic programming. However, finding optimal policies for these problems is computationally hard in general and thus, heuristic solutions are employed in practice. Deep learning can be used as a tool for designing better heuristics in such problems. In this paper, the problem of active sequential hypothesis testing is considered. The goal is to design a policy that can reliably infer the true hypothesis using as few samples as possible by adaptively selecting appropriate queries. This problem can be modeled as a POMDP and bounds on its value function exist in literature. However, optimal policies have not been identified and various heuristics are used. In this paper, two new heuristics are proposed: one based on deep reinforcement learning and another based on a KL-divergence zero-sum game. These heuristics are compared with state-of-the-art solutions and it is demonstrated using numerical experiments that the proposed heuristics can achieve significantly better performance than existing methods in some scenarios.
\end{abstract}

\section{Introduction}
Information theory provides us with a quantitative framework \cite{shannon2001mathematical, cover2012elements} to analyze various notions associated with information processing such as communication and storage. Some of the major advances in data storage and communication have been facilitated by information theory. Information theory has also been very successful in identifying performance limits such as channel capacity and compression rate. It guarantees the existence of policies that achieve optimal performance but in many cases, finding these optimal policies (like capacity-achieving encoding and decoding schemes) can be a difficult task. Many problems in statistics, such as hypothesis testing, also have strong connections with information theory.

Stochastic control theory \cite{kumar2015stochastic,bertsekas2005dynamic} provides us with a framework to analyze sequential decision-making problems under uncertainty. Many real world decision-making problems can be modeled as Markov Decision Processes (MDPs) or Partially Observable Markov Decision Processes (POMDPs). It has been widely used in the areas of artificial intelligence, robotics and finance. Stochastic control theory provides us with strong tools such as Dynamic Programming (DP) that can help us characterize optimal solutions for these decision problems. For instance, optimal solutions for MDPs with complete model information can be computed efficiently using dynamic programming \cite{bertsekas2005dynamic}. This efficiency, however, does not extend to POMDPs. It is known that finding optimal solutions for POMDPs in general is a PSPACE-hard problem \cite{papadimitriou1987complexity}.

Because of the computational hardness of these problem, various heuristic solutions are employed in practice. For example, Point Based Value Iteration \cite{pineau2003point} is a well-known heuristic for POMDPs. In this work, we examine the possibility of using deep learning to design better heuristics for problems in information and control theory. Deep learning is an emerging branch of machine learning and has found tremendous success in the areas of image and text processing \cite{lecun2015deep}. Deep neural networks are universal approximators \cite{hornik1989multilayer} and these networks can be trained in a supervised manner using the backpropagation algorithm \cite{hagan1994training}. Deep neural networks have lately been used to solve problems in communication \cite{farsad2017detection,farsad2018deep} and reinforcement learning \cite{mnih2015human}.

In this work, we consider the problem of active sequential hypothesis testing, which involves a combination of information and control theory. The aim of active hypothesis testing is to infer an unknown hypothesis based on observations. The agent can adaptively make queries to obtain observations and we seek to design a sequential query selection policy that can reliably infer the underlying hypothesis using few queries. We define a notion of confidence and reformulate this problem as a confidence maximization problem in a fixed sample-size setting. The asymptotic version of the confidence maximization problem can be seen as an infinite-horizon, average-reward MDP.

We design heuristics for this confidence maximization problem using deep neural networks. We first examine a design framework based on Recurrent Neural Networks (RNNs). RNNs are a category of neural networks with a recurring neural unit which maintains a hidden state vector for each input instance. They have been used for solving sequential problems with success \cite{sutskever2014sequence, cho2014learning}. Their ability to store long-term dependencies of sequences within hidden states makes them apt for the task. One of the most popular and successful variants of recurrent networks are Long-Short Term Memory (LSTM) networks \cite{hochreiter1997long} which maintain their state using forget, input and output gates. The underlying structure of recurrent networks fits naturally for active hypothesis testing. We explore an LSTM architecture that can be trained simultaneously to adaptively select queries as well as learn to infer the true hypothesis based on observations obtained. We observe that the model manages to learn to infer the hypothesis for a given set of observations but fails to learn the query selection policy. We discuss the details of this architecture in Appendix \ref{rnn}.

We then design a heuristic based on deep reinforcement learning \cite{mnih2015human}. In this heuristic, the agent simulates the MDP associated with the confidence maximization problem. Based on its simulated experience, the agent tries to learn the optimal query selection policy using deep reinforcement learning. We observe in our numerical experiments that this heuristic policy comes very close to optimality. The details of this approach are discussed in Section \ref{dqndes}. In addition to the neural network based heuristics, we introduce a heuristic based on a KL-divergence zero-sum game. This policy is adaptive and also achieves near-optimal performance in our numerical experiments. The details of the policy are discussed in Section \ref{kzg}.

The rest of the paper is organized as follows. In Section \ref{prior}, we describe the prior works related to active hypothesis testing and deep learning. In Section \ref{notation}, we discuss the mathematical notation used in this paper. The confidence maximization problem is formulated in Section \ref{prob} and expressed as an MDP in Section \ref{mdpform}. The deep reinforcement learning approach for active hypothesis testing is discussed in Section \ref{dqndes}. In Section \ref{numerical} we describe a few policies from prior works and compare them with our designed policies based on numerical experiments. We conclude the paper in Section \ref{conc}.

\subsection{Related Work}\label{prior}
Active hypothesis testing was first formulated by Chernoff in \cite{chernoff1959sequential} inspired by Wald's Sequential Probability Ratio Test (SPRT) \cite{wald1973sequential}. Thereafter, this work has been extended in various ways. In \cite{nitinawarat2013controlled}, the problem of multihypothesis testing is considered in both fixed sample size and sequential settings. In \cite{naghshvar2013active}, a Bayesian setting is examined with a random stopping time and is formulated as a POMDP. Upper and lower bounds on the optimal value function of this POMDP were derived and some heuristic policies were proposed. All these works provide heuristic policies that are asymptotically optimal. However, optimal policies for the non-asymptotic formulations are not known. Furthermore, most of the heuristics proposed in these works are almost open-loop and randomized policies. This motivates us to seek better heuristics.

The idea of using deep neural networks to solve POMDPs is relatively less explored. Reinforcement learning usually assumes perfect state observability. In \cite{hausknecht2015deep, karkus2017qmdp}, the authors aim to perform deep reinforcement learning under partial observability. They use a combination of convolutional neural networks and recurrent neural networks to achieve this. In this model, the agent does not have model information and thus, cannot directly make Bayesian belief updates. The network model in \cite{egorov2015deep} is very similar to our deep Q-network. However, the model in \cite{egorov2015deep} cannot be used directly for hypothesis testing due to some issues discussed in Section \ref{issues}. We make appropriate modifications to rectify these issues. To the best of our knowledge, deep neural networks have not been used in the context of active hypothesis testing and our neural network design framework is the first of its kind.

\subsection{Notation}\label{notation}
Random variables/vectors are denoted by upper case boldface letters, their realization by the corresponding lower case letter. We use calligraphic fonts to denote sets (e.g. $\mathcal{U}$) and $\Delta \mathcal{U}$ is the probability simplex over a finite set $\mathcal{U}$. In general, subscripts are used as time indices. There is an exception (${\rho}_j(n)$) to this convention where the subscript denotes the hypothesis and $n$ denotes time. For time indices $n_1\leq n_2$, $\rv{X}_{n_1:n_2}$ is the abbreviated notation for the variables $(\rv{X}_{n_1},\rv{X}_{n_1+1},...,\rv{X}_{n_2})$.
For a strategy $g$, we use $\Py^g[\cdot]$ and $\E^g[\cdot]$ to indicate that the probability and expectation depend on the choice of $g$. The Shannon entropy of a discrete distribution $p$ over a space $\mathcal{Y}$ is given by
\begin{equation}
H(p) = -\sum_{y \in \mathcal{Y}}p(y)\log p(y).
\end{equation}
The Kullback-Leibler divergence between distributions $p$ and $q$ is given by
\begin{equation}
D(p || q) = \sum_{y \in \mathcal{Y}}p(y)\log\frac{p(y)}{q(y)}.
\end{equation}

\section{Problem Formulation}\label{prob}

%\subsection{Notation}

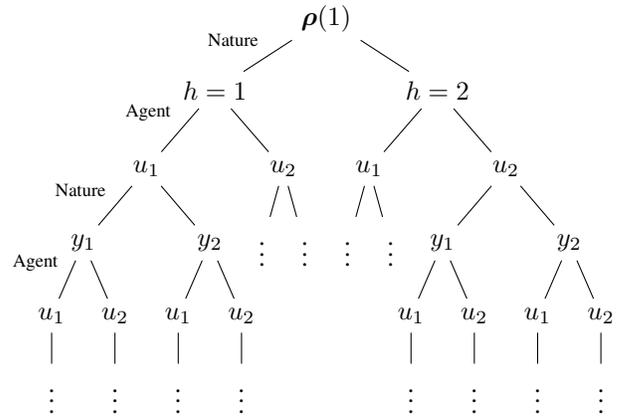
\begin{figure}
\centering
\scalebox{1}{
\begin{forest}
[$\vct{\rho}(1)$
[${h=1}$, edge label={node[midway,above left,font=\scriptsize]{Nature}} 
  %[$\varnothing$,edge label={node[midway,above left,font=\scriptsize]{Agent 1}}]
  [$u_1$,edge label={node[midway,above left,font=\scriptsize]{Agent}} 
   [$y_1$,edge label={node[midway,above left,font=\scriptsize]{Nature}}
   	%[$\varnothing$,edge label={node[midway,above left,font=\scriptsize]{Agent 1}}]
   	[$u_1$,edge label={node[midway,above left,font=\scriptsize]{Agent}}[$\vdots$]]
	[$u_2$[$\vdots$]]]
   [$y_2$
   	%[$\varnothing$]
   	[$u_1$ [$\vdots$]]
	[$u_2$[$\vdots$]]]
  ]
  [$u_2$
  [$\vdots$] [$\vdots$]]
]
[${h=2}$ 
  %[$\varnothing$,edge label={node[midway,above left,font=\scriptsize]{Agent 1}}]
  [$u_1$
  [$\vdots$] [$\vdots$]]
    [$u_2$, 
   [$y_1$,
   	%[$\varnothing$,edge label={node[midway,above left,font=\scriptsize]{Agent 1}}]
   	[$u_1$,[$\vdots$]]
	[$u_2$[$\vdots$]]]
   [$y_2$
   	%[$\varnothing$]
   	[$u_1$ [$\vdots$]]
	[$u_2$[$\vdots$]]]
  ]
]
]
\end{forest}}
\caption{Agent's choices and subsequent observations represented as a tree. Every instance of the probability space can be uniquely represented by a path in this tree.}
\label{fig:illustree}
\end{figure}
Let $\mathcal{H} \subset \N$ be a finite set of hypotheses and let $\rv{H}$ be the true hypothesis. At each time $n \in \N$, the agent can perform an experiment $\rv{U }_n \in \mathcal{U}$ and obtain an observation $\rv{Y}_n \in \mathcal{Y}$. The relation between $\rv{U}_n$ and $\rv{Y}_n$ is given by
\begin{equation}
\rv{Y}_n = \xi(\rv{H}, \rv{U}_n,\rv{W}_n),
\end{equation}
where $\rv{W}_n$ is a collection of independent primitive random variables. Thus, all the observations are independent conditioned on the hypothesis and the experiment. The probability of observing $y$ after performing an experiment $u$ under hypothesis $h$ is denoted by $p_h^u(y)$. For simplicity, let us also assume that the sets $\mathcal{U}$ and $\mathcal{Y}$ are finite.

The information available at the agent at time $n$ is
\begin{equation}
\rv{I}_n = \{\rv{U}_{1:n-1},\rv{Y}_{1:n-1}\}.
\end{equation}
Actions of the agent at time $n$ can be functions of $\rv{I}_n$ (see Fig. \ref{fig:illustree}). Let the experiment selection policy be
\begin{equation}
\rv{U}_n = g_n(\rv{I}_n).
\end{equation}
The sequence of all the policies $\{g_n\}$ is denoted by $g$ which is referred to as a \emph{strategy}. Let the collection of all such strategies be $\mathcal{G}$. Using the available information, the agent forms a \emph{posterior belief} $\vct{\rho}(n)$ on $\rv{H}$ at time $n$ which is given by
\begin{equation}
\rho_h(n) = \Py[\rv{H} = h \mid \rv{Y}_{1:n-1},\rv{U}_{1:n-1}].
\end{equation}

\begin{figure}[]

\hspace{0.0in}
\begin{tikzpicture}
\begin{axis}[xlabel = $p$, xmajorgrids = true,ymajorgrids = true,grid style = dashed,legend pos=north west]
\addplot[domain = 0.0001:0.9999,samples = 1000]{ln(x/(1-x))};
\legend{$\log\frac{p}{1-p}$}
\end{axis}
\end{tikzpicture}
\caption{The logit function is the inverse of the logistic sigmoid function $1/(1+e^{-x})$. It is widely used in statistics and machine learning to quantify confidence level \cite{hosmer2013applied}.}
\label{logitplot}
\end{figure}
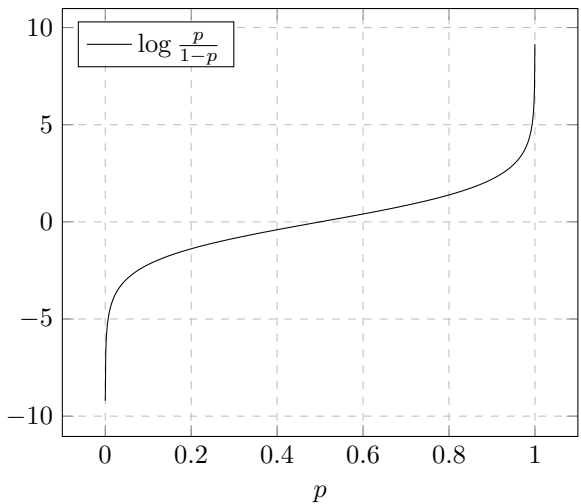

\begin{definition}[Bayesian Log-Likelihood Ratio]
The Bayesian log-likelihood ratio $\mathcal{C}_h(\vct{\rho})$ associated with an hypothesis $h \in \mathcal{H}$ is defined as
\begin{equation}
\mathcal{C}_h(\vct{\rho}) := \log\frac{\rho_h}{1-\rho_h}.\\
\end{equation}
\end{definition}
\vspace{0.05in}
The Bayesian log-likelihood ratio (BLLR) is the logarithm of the ratio of the probability that hypothesis $h$ is true versus the probability that hypothesis $h$ is not true. BLLR can be interpreted as a \emph{confidence level} on hypothesis $h$ being true in logit form, which is also referred to as \emph{log-odds} in statistics \cite{hosmer2013applied}. The logit function is the inverse of the logistic sigmoid function. Notice that the posterior belief $\rho_h$ and BLLR are related by the bijective increasing logit function (See Fig. \ref{logitplot}).

The objective is to design an experiment selection strategy $g$ for the agent such that the confidence level $\mathcal{C}_{\rv{H}}$ on the true hypothesis $\rv{H}$ increases as quickly as possible. In other words, the total reward after acquiring $N$ observations is the average rate of increase in the confidence level on the true hypothesis $\rv{H}$ and is given by
\begin{equation}
\frac{\mathcal{C}_{\rv{H}}(\vct{\rho}(N+1))-\mathcal{C}_{\rv{H}}(\vct{\rho}(1))}{N}.
\end{equation}
More explicitly, we seek to design a policy $g$ that maximizes the asymptotic expected reward $R(g)$ which is defined as
\begin{align*}
R(g) &:= \lim_{N \to \infty} \inf \frac{1}{N} \;\E^g \left[\mathcal{C}_{\rv{H}}(\vct{\rho}(N+1))- \mathcal{C}_{\rv{H}}(\vct{\rho}(1)) \right].
\end{align*}
Since the initial confidence $\mathcal{C}_{\rv{H}}(\vct{\rho}(1))$ is a constant, we can ignore it for large values of $N$. Henceforth, we refer to this problem as the \emph{Expected Confidence Maximization} (ECM) problem.

\begin{remark}
Generally, the objective is to maximize the decay rate of Bayesian error probability \cite{nitinawarat2013controlled}, or to use a stopping time and optimize a linear combination of expected stopping time and expected error probability \cite{nitinawarat2013controlled,naghshvar2013active}. Our problem formulation is mathematically different from these frameworks but conceptually, all the formulations aim to capture the same phenomenon which is to infer the true hypothesis quickly and reliably. The precise mathematical relationship between these formulations is yet to be understood and is an avenue for future work.
\end{remark}

To describe an upper bound on the optimal performance of the confidence maximization problem, we state the following theorem without proof.
\begin{theorem}For any query selection policy $g$ and any hypothesis $h$, we have
\begin{align}
 \lim_{N \to \infty} \sup \frac{1}{N} \;\E^g \left[\mathcal{C}_{\rv{H}}(\vct{\rho}(N+1)) \mid \rv{H} = h \right]  \leq R^*_h,
\end{align}
where
\begin{align}
R_i^* := \max_{\vct{\alpha} \in \Delta\mathcal{U}} \min_{j\neq i} \sum_{u}\alpha_u D(p_i^u || p_j^u).
\end{align}
Further,  if the underlying hypothesis $\rv{H} = h$, we have
\begin{align}
\lim_{N \to \infty} \sup \frac{1}{N}  \left[\mathcal{C}_{{h}}(\vct{\rho}(N+1)) \right]  \leq R^*_h,
\end{align}
with probability 1.
\end{theorem}

The upper bound on the expected confidence rate can be obtained using dynamic programming for infinite-horizon, average reward MDPs and the same inequality in an almost sure sense can be obtained with the help of Strong Law of Large Numbers (SLLN).

%Agent 1 needs to minimize the error probability of Agent 2's decision as quickly as possible. Thus, the objective is to design a strategy $g$ for Agent 1 that maximizes the following reward
%\begin{align}
%\nonumber K(g) &= \E \left[\lim_{N \to \infty} \inf \frac{1}{N} \;\E^g \left[\log \frac{1-\rho_{h}(1)}{1-\rho_{h}(N+1)}\mid \rv{H} \right]\right]\\
%&=: \E [J(g,\rv{H})].
%\end{align}
%The reward $K(g)$ is the expected average decay rate of error probability and similarly, $J(g,h)$ is the conditional expected average decay rate of error probability. Since Agent 1 can use the knowledge of $\rv{H}$, we can design the strategy $g$ separately for each $h \in \mathcal{H}$. Let the strategy corresponding to hypothesis $h$ be $g^h$.

\section{Markov Decision Process Formulation}\label{mdpform}
In this section, we show that the problem of maximizing $R(g)$ can be formulated as an infinite-horizon, average-cost MDP problem. The state of the MDP is the posterior belief $\vct{\rho}(n)$. The agent's observation and action spaces are the same as in Section \ref{prob}.
The posterior belief is updated using Bayes' rule. Thus, if $\rv{U}_n = u$ and $\rv{Y}_n = y$, we have
\begin{align}
{\rho}_{h}(n+1) = \frac{\rho_h(n)p_h^u(y)}{\sum_{h'}\rho_{h'}(n)p^u_{h'}(y)}.\label{bayes}
\end{align} 
For convenience, let us denote the Bayes' update in (\ref{bayes}) by
\begin{align}
\vct{\rho}({n+1}) &= F(\vct{\rho}(n),\rv{U}_n,\rv{Y}_{n}).
\end{align}
Thus, we have
\begin{align*}
\Py[\vct{\rho}&(n+1) = F(\vct{\rho}(n),u,y) \mid \rv{I}_{n}, \rv{U}_n = u] \\
%&= \Py[\rv{Y}_n = y \mid \rv{I}_n,\rv{U}_n = u]\\
%&= \sum_{h \in \mathcal{H}}\Py[\rv{H} = h \mid  \rv{I}_n,\rv{U}_n = u]\Py[\rv{Y}_n = y \mid \rv{I}_n,\rv{U}_n = u,\rv{H} = h]\\
&=  \Py[\rv{Y}_n = y \mid \rv{I}_n,\rv{U}_n = u] = \sum_{h \in \mathcal{H}}\rho_h(n)p^u_h(y).
\end{align*}
Clearly, the dynamics of this system are controlled Markovian. The expectation of the average confidence rate under a strategy ${g}$ is given by
\begin{align}
\label{objfin}R_N(g) &\doteq\frac{1}{N}\E^{{g}} \left[\mathcal{C}_{\rv{H}}(\vct{\rho}(N+1))- \mathcal{C}_{\rv{H}}(\vct{\rho}(1)) \right]\\
&= \frac{1}{N}\E^{{g}} \sum_{n = 1}^N \left[\mathcal{C}_{\rv{H}}(\vct{\rho}(n+1))- \mathcal{C}_{\rv{H}}(\vct{\rho}(n)) \right]\\
%\nonumber&= \frac{1}{N}\E^{{g}} \sum_{n = 1}^N \E \left[\mathcal{C}_{\rv{H}}(\vct{\rho}(n+1))- \mathcal{C}_{\rv{H}}(\vct{\rho}(n)) \mid \rv{I}_n,\rv{U}_n,\rv{Y}_n\right]\\
%\nonumber &= \frac{1}{N}\E^{{g}} \sum_{n = 1}^N \E \left[\mathcal{C}_{\rv{H}}(\vct{\rho}(n+1))- \mathcal{C}_{\rv{H}}(\vct{\rho}(n)) \mid \vct{\rho}(n),\rv{U}_n,\rv{Y}_n\right]\\
&=: \frac{1}{N}\E^{{g}} \sum_{n = 1}^N r(\vct{\rho}(n),\rv{U}_n,\rv{Y}_n).
\end{align}
Thus, the instantaneous reward for this MDP is $r(\vct{\rho},u,y)$, i.e. if the state is $\vct{\rho}$, the experiment performed is $u$ and the observation is $y$, then the instantaneous reward is given by
\begin{align}
r(\vct{\rho},u,y) &= \mathcal{C}(F(\vct{\rho},u,y)) - \mathcal{C}(\vct{\rho}),
\end{align}
where for any belief state $\vct{\rho}$
\begin{align}
\mathcal{C}(\vct{\rho}) &= \sum_{i \in \mathcal{H}}\rho_i\log\frac{\rho_i}{1-\rho_i} = \sum_{i \in \mathcal{H}}\rho_i\mathcal{C}_i(\vct{\rho}).
\end{align}
We refer to the function $\mathcal{C}(\vct{\rho})$ as \emph{Average Bayesian Log-Likelihood Ratio} (ABLLR). Note that this is almost identical to the notion of average log-likelihood ratio $U(\vct{\rho}) = -\mathcal{C}(\vct{\rho})$ in \cite{naghshvar2013active1}, which was used to design a greedy heuristic for the active hypothesis testing problem. 

The objective in this MDP problem is to find a strategy $g^*$ that maximizes the following average expected reward
\begin{align}
R(g) = \lim_{N \to \infty}\inf\frac{1}{N}\sum_{n=1}^N \E^g(r(\vct{\rho}(n),\rv{U}_n,\rv{Y}_n)).
\end{align}

\section{Deep Q-learning for Hypothesis Testing}\label{dqndes}
In this section, we describe our deep learning approach for policy design for active sequential hypothesis testing. We use a variant of the Deep Q-Network (DQN) introduced in \cite{mnih2015human}, which is a learning agent that combines reinforcement learning with deep neural networks. It is an adaptation of a popular off-policy Temporal Difference (TD) learning algorithm, known as Q-learning \cite{sutton}. 

\begin{figure}[]

\centering
\includegraphics[width=0.8\columnwidth]{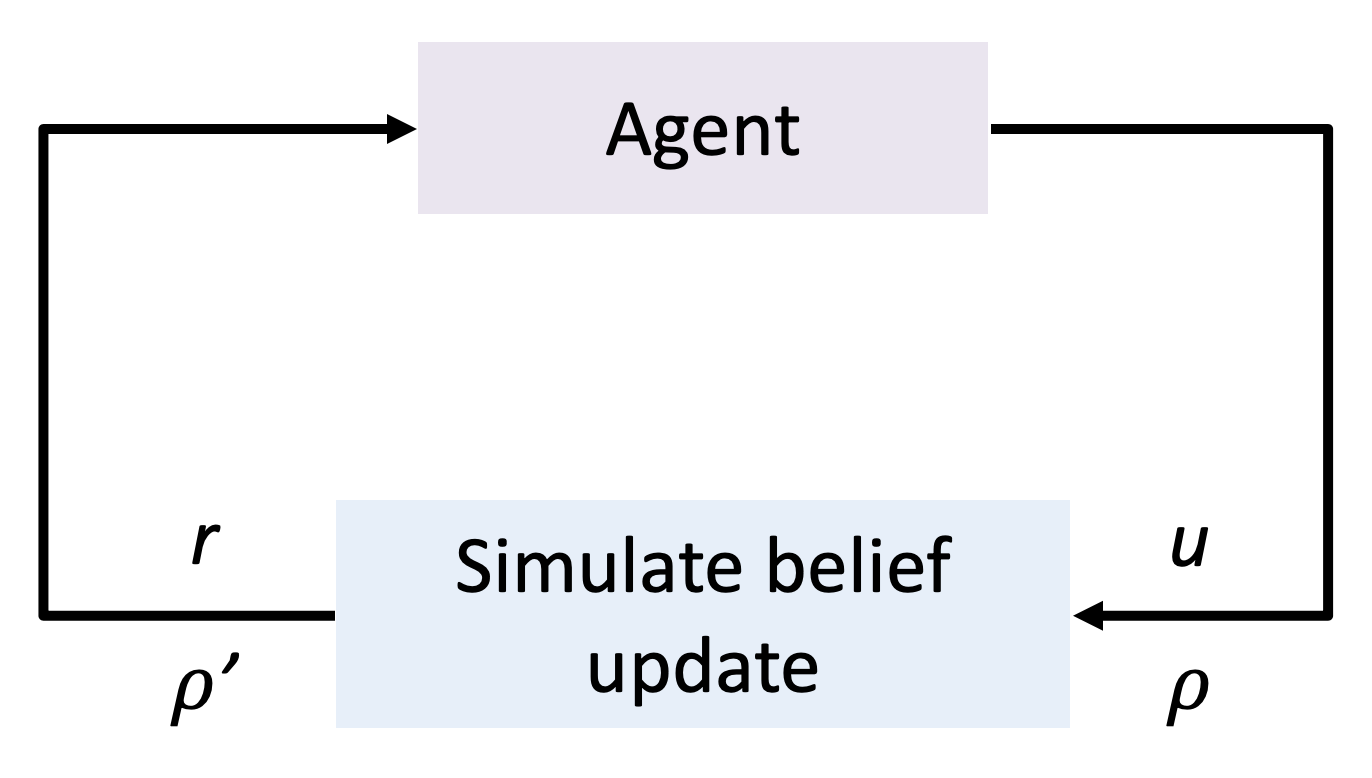}%

\caption{The agent performs a query $u$ at some state $\vct{\rho}$. The environments simulates the belief update using $u$ and $\vct{\rho}$ to generate the update belief $\vct{\rho}'$ and its associated reward $r$.}
\label{agent}
\end{figure}

We create an artificial environment that simulates the Bayesian belief update (\ref{bayes}) over multiple \emph{episodes}. The duration ($N$) of each episode is fixed. At the beginning of each episode, the underlying hypothesis $\rv{H}$ is randomly selected with probability $\vct{\rho}(1)$ and it remains fixed over the episode's duration. At any given time, the agent interacts with this environment by making a query ($u$) based on the current state ($\vct{\rho}$) using an appropriate exploration strategy (e.g. $\epsilon$-greedy exploration \cite{sutton}). The environment then reveals the next state ($\vct{\rho}'$) and its associated reward ($r$) to the agent. We refer to $(\vct{\rho},u,\vct{\rho}',r)$ as an \emph{experience tuple}. Using this information, the agent updates its target policy $g$. This iterative simulated learning process, schematically illustrated in Figure \ref{agent}, is repeated until a convergence criterion is met. We elucidate this methodology in greater detail in the following sub-sections.
%
%In short, the DQN maintains an action-value function $Q(s,a)$, where $s$ and $a$ denote the state and action, respectively. At time $n$, the learning agent interacts with the environment by performing an action $a_n$ and observes the state transition from $s_n$ to $s_{n+1}$ as well as the corresponding reward $r_n$. Using the tuple $(s_n,a_n,s_{n+1},r_n)$, the action-value function is updated accordingly and this process is repeated until the $Q$ function converges.
%
%When the state and action spaces are finite, the $Q$ function can simply be represented as a matrix.
%
\subsection{Discounted Reward Formulation}
The Q-learning algorithm is designed for a discounted reward MDP formulation. Therefore, we first convert our average reward formulation in Section \ref{mdpform} to a discounted reward formulation with a discount factor $\gamma < 1$. For an experiment selection policy $g$, let
\begin{align}
R^d(g) := \E^g\left[\sum_{n=1}^\infty\gamma^{n-1}r(\vct{\rho}(n),\rv{U}_n,\rv{Y}_n)\right],
\end{align}
be the total discounted reward. Since $r(\vct{\rho},u,y)$ is uniformly bounded, the discounted reward $R^d(g)$ is also bounded and well-defined for any policy $g$. Our objective now is to find a strategy $g^*$ that maximizes $R^d(g)$.
\begin{remark}
When the state and action spaces of an MDP are finite, it is well-known \cite{bertsekas2005dynamic} that the discounted reward and the average reward formulations are equivalent for a sufficiently large discount factor $\gamma$. However, the state space herein is uncountably infinite and this equivalence may not necessarily hold. Nonetheless, we observe in our numerical experiments that the solution to the discounted reward formulation is near-optimal with respect to the average reward formulation.
\end{remark}

\subsection{Action-value Function}
The action-value function \cite{sutton} for a policy $g$ is defined as
\begin{align}
q_g(\vct{\rho}, u) := \E^g\left[\sum_{k=0}^{\infty} \gamma^k\rv{r}_{k+1} \mid \vct{\rho}(1) = \vct{\rho}, \rv{U}_1 = u\right],
\end{align}
where $\rv{r}_n = r(\vct{\rho}(n),\rv{U}_n,\rv{Y}_n)$. Let $g^*$ be an optimal policy with respect to the discounted reward formulation and let $q^*(\vct{\rho}, u)$ be its corresponding action-value function. Then the optimal action-value function satisfies the fixed point equation, also known as Bellman optimality equation \cite{sutton},
\begin{align}
q^*(\vct{\rho},u) = \E[r(\vct{\rho},u,\rv{Y}) + \max_{u'}q^*( F(\vct{\rho},u,\rv{Y}),u')],
\end{align}
for every belief state $\vct{\rho}$ and query $u$. Note that the source of randomness in the fixed point equation is the variable $\rv{Y}$ and the expectation is with respect to the distribution $\sum_{h \in \mathcal{H}}\rho_hp^u_h(y)$. Further, if a policy $g$ is such that, for every belief state $\vct{\rho}$,
\begin{equation}
g(\vct{\rho}) = \arg\max_{u}q^*(\vct{\rho},u),
\end{equation} 
then $g$ is an optimal policy with respect to the discounted reward formulation. Thus, finding an optimal policy $g^*$ can be reduced to finding the optimal action-value function $q^*$. The optimal action-value function can be obtained using the Q-learning algorithm in \cite{sutton} when the state and action spaces are finite. However, the state space is infinite in our case and thus, we need a different approach to find the optimal action-value function.

\subsection{Action-value Function as a Deep Neural Network}
\begin{figure}[]

\centering
\includegraphics[width=0.8\columnwidth]{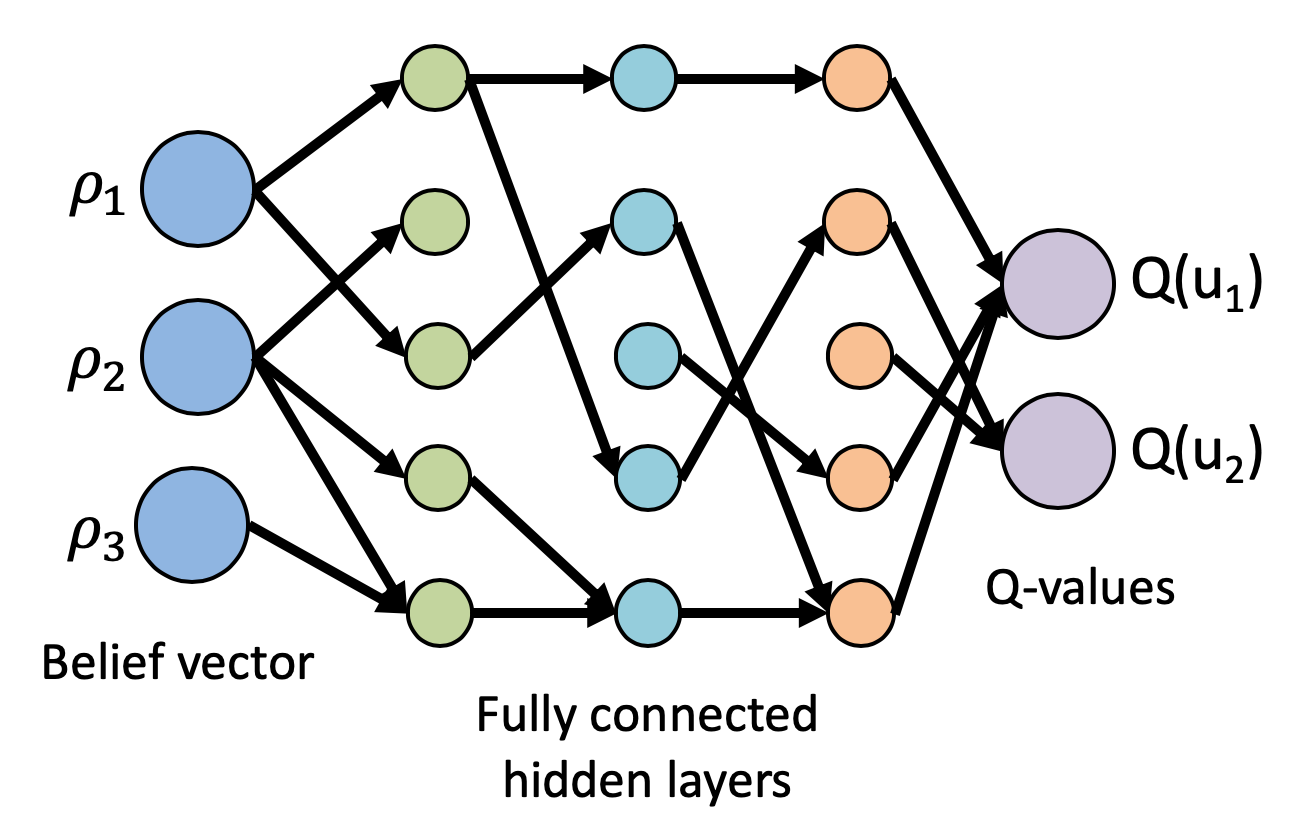}%

\caption{The neural network takes the belief vector $\vct{\rho}$ as the input and outputs the Q-values for each action $u$. The hidden layers are fully connected with non-linear activation. Only the final layer has linear activation.}
\label{nnpic}
\end{figure}

The first challenge in performing Q-learning with an infinite state space is to find an appropriate representation for the action-value function. Notice that the posterior belief $\vct{\rho}$ is a finite-dimensional vector and the action space is finite. We can thus represent the action-value function as a deep neural network which takes posterior belief $\vct{\rho}$ as an input and outputs the action-value vector of dimension $|\mathcal{U}|$ as illustrated in Figure \ref{nnpic}. The neural network is parameterized by a finite collection of weights $\theta$ and henceforth, we refer to the output of this neural network as $Q_\theta(\vct{\rho},u)$.

The second challenge lies in making the Q-learning updates. We would ideally like to make an update of the following form
%\begin{align*}
%Q(\vct{\rho}(n),\rv{U}_n) \leftarrow &\; Q(\vct{\rho}(n),\rv{U}_n) + \alpha[r_n \\
%&+ \gamma\max_{u}Q(\vct{\rho}(n+1),u) - Q(\vct{\rho}(n),\rv{U}_n)].
%\end{align*}
\begin{align*}
Q'(\vct{\rho},u) \leftarrow \; Q(\vct{\rho},u) + \zeta[r &+ \gamma\max_{u'}Q(\vct{\rho}',u') - Q(\vct{\rho},u)],
\end{align*}
where $(\vct{\rho},u,\vct{\rho}',r)$ is an \emph{experience} tuple. Notice, however, that the action-value function is characterized by a collection of weights ($\theta$) and thus, one has to update these weights so that the neural network outputs the corresponding updated Q-values $Q'(\vct{\rho},u)$. To achieve this, we can modify the weights $\theta$ using gradient descent such that the following Mean-Squared Error (MSE) loss is minimized
\begin{align}
L(\theta) = (Q'(\vct{\rho},u) - Q_\theta(\vct{\rho},u))^2.
\end{align}
This naive update rule can make the network unstable because it closely fits the network to the updated Q-value $Q'(\vct{\rho},u)$ for the current state-action pair $(\vct{\rho},u)$ but the Q-values associated with other state-action pairs may be disturbed. To ensure this does not happen, a method known as experience replay \cite{mnih2015human} is employed. At each time, the agent stores experience tuples $(\vct{\rho},u,\vct{\rho}',r)$ in its memory $\mathscr{D}$. Whenever the weights are updated, a random mini-batch $\mathscr{B}$ of experience tuples is sampled from the memory and the MSE loss is minimized using gradient descent over this mini-batch with the following loss function
\begin{align}
L(\theta) = \sum_{(\vct{\rho},u,\vct{\rho}',r) \in \mathscr{B}}(Q'(\vct{\rho},u) - Q_\theta(\vct{\rho},u))^2.
\end{align}

\subsection{Additional Challenges}\label{issues}
Generally, it is necessary to explore all the states to learn the state transition and reward structure. However, since the state space is uncountably infinite, we cannot possibly explore all the states. Therefore, we choose a large value for $\epsilon$ ($\approx 0.8$)  so that the state space is sufficiently explored. We observe in our numerical experiments that training over a large number of episodes results in an efficient query selection policy despite this exploration issue.

Another challenge is that as the belief on the true hypothesis gets close to 1, the belief on all the alternate hypotheses becomes very small. Improvement in the confidence level, i.e. the instantaneous reward $r$ is very sensitive to the belief on alternate hypotheses. Thus, the DQN fails to select optimal queries when the belief on alternate hypotheses is too small. To counter this, we normalize the belief on the alternate set of hypotheses and augment it to the belief vector. The normalized alternate belief is denoted by $\tilde{\vct{\rho}}$ and is given by
\begin{align}
\tilde{\rho}_j = \frac{\rho_j}{1-\rho_i},
\end{align}
where $i$ is the most likely hypothesis with respect to the belief $\vct{\rho}$ and $j \neq i$.

The overall Deep Q-learning algorithm for active sequential hypothesis testing is described in Algorithm \ref{mainalgo}. The agent and the environment operate in an interleaved manner. Their combined behavior is captured by Algorithm \ref{mainalgo}. The comment on each instruction specifies whether the instruction is meant for the agent (A) or the artificial environment (E). The Q-value update is denoted by $\mathrm{QUP}_\theta$ and is given by
\begin{align*}
\mathrm{QUP}_\theta&(\vct{\rho},u,\vct{\rho}',r) \\
&= Q_\theta(\vct{\rho},u) + \zeta[r + \gamma\max_{u'}Q_\theta(\vct{\rho}',u') - Q_\theta(\vct{\rho},u)].
\end{align*}
Note that $\zeta$ is a small constant less than 1. Note that the letter $\alpha$ is generally used in place of $\zeta$. We select $\zeta$ to avoid notational conflict with distribution $\vct{\alpha}_i^*$ which will be introduced later.

\begin{algorithm}
\caption{Deep Q-learning algorithm for active sequential hypothesis testing}
\begin{algorithmic}[1]
\State Initialize memory $\mathscr{D}$ to capacity $K$  \Comment{A}
\State Initialize DQN with random weights $\theta$  \Comment{A}
\For{episode = 1, EpiNum}
\State Randomly select $\rv{H}$ with prob. $\vct{\rho}(1)$ \Comment{E}
\State Initialize state $\vct{\rho} = \vct{\rho}(1)$  \Comment{A}
\For{$n = 1,N$}
\State With probility $\epsilon$, select random query $u$  \Comment{A}
\State Otherwise, select $u = \arg\max_uQ_\theta(\vct{\rho},u)$  \Comment{A}
\State Perform query $u$  \Comment{A}
\State Generate $\rv{Y} =  \xi(\rv{H}, u,\rv{W})$  \Comment{E}
\State Update belief $\vct{\rho}' = F(\vct{\rho},u,\rv{Y})$  \Comment{E}
\State Compute reward $r = \mathcal{C}(\vct{\rho}') - \mathcal{C}(\vct{\rho})$  \Comment{E}
\State Reveal $\vct{\rho}'$ and $r$ to A  \Comment{E}
\State Store $(\vct{\rho},u,\vct{\rho}',r)$ in $\mathscr{D}$ \Comment{A}
\State Assign $\vct{\rho} \leftarrow \vct{\rho}'$ \Comment{A}
\State Sample random minibatch $\mathscr{B}$ from $\mathscr{D}$ \Comment{A}
\State Duplicate DQN $\theta' \leftarrow \theta$ \Comment{A}
\For{epoch = 1, EpochNum}
\For{each $(\hat{\vct{\rho}},\hat{u},\hat{\vct{\rho}}',\hat{r})$ in $\mathscr{B}$}
\State $Q'(\hat{\vct{\rho}},\hat{u}) \leftarrow \mathrm{QUP}_\theta(\hat{\vct{\rho}},\hat{u},\hat{\vct{\rho}}',\hat{r})$ \Comment{A}
\State Perform gradient descent step on \Comment{A}
\State $(Q'(\hat{\vct{\rho}},\hat{u}) - Q_{\theta'}(\hat{\vct{\rho}},\hat{u}))^2$
\EndFor
\EndFor
\State Assign $\theta \leftarrow \theta'$ \Comment{A}
\EndFor
\EndFor
\State \textbf{return} DQN $\theta$

\end{algorithmic}
\label{mainalgo}
\end{algorithm}

\section{Numerical Experiments}\label{numerical}
In this section, we numerically compare our DQN model with other popular heuristics used for active hypothesis testing. We also propose a new heuristic based on a Kullback-Leibler divergence zero-sum game and demonstrate numerically that this heuristic's performance is close to the maximum achievable confidence rate. We first briefly describe all the heuristics we use in our experiments.

\subsection{Extrinsic Jensen-Shannon (EJS) Divergence}
Extrinsic Jensen-Shannon divergence as a notion of information was first introduced in \cite{naghshvar2012extrinsic}. Using our notation, EJS for a query $u$ at some belief state $\vct{\rho}$ is simply the expected instantaneous reward, i.e.
\begin{align}
EJS(\vct{\rho},u) = \E [\mathcal{C}(F(\vct{\rho},u,\rv{Y})) - \mathcal{C}(\vct{\rho})] .
\end{align}
Notice that the only random variable in the expression above is $\rv{Y}$ and the expectation is with respect to the distribution $\sum_{h \in \mathcal{H}}\rho_hp^u_h(y)$ on $\mathcal{Y}$. The EJS heuristic selects the experiment $u$ that maximizes $EJS(\vct{\rho},u)$ for a given state $\vct{\rho}$.

\subsection{Open Loop Verification (OPE)}
Open loop verification policy is the most widely used policy in prior literature \cite{naghshvar2013active,nitinawarat2013controlled}. In this heuristic, the agent first explores for a while using an appropriate exploration strategy. Whenever the confidence on some hypothesis $i$ is large enough, i.e. $\rho_i > \bar{\rho}$, the queries are randomly selected in an open-loop manner from the distribution $\vct{\alpha}_i^*$ which is defined as
\begin{align}
\vct{\alpha}_i^* &:= \arg \max_{\vct{\alpha} \in \Delta\mathcal{U}} \min_{j\neq i} \sum_{u}\alpha_u D(p_i^u || p_j^u),
\end{align}
where $\Delta\mathcal{U}$ is the set of all distributions over the set of experiments $\mathcal{U}$. We refer to this phase as the \emph{verification phase}. In our implementation, we use EJS for the exploration phase and set the threshold $\bar{\rho} = 0.7$.

\subsection{KL-divergence Zero-sum Game (HEU)}\label{kzg}
This heuristic is similar to OPE but the query selection policy in the verification phase is adaptive. For each hypothesis $i$, we can formulate a zero-sum game \cite{osborne1994course} in which the first player (maximizing) selects an experiment $u \in \mathcal{U}$ and the second player (minimizing) selects an alternate hypothesis $j \in \tilde{\mathcal{H}}_i := \mathcal{H}\setminus \{i\}$. The payoff for this zero-sum game is the KL-divergence $D(p_i^u || p_j^u)$. Whenever $\rho_i > \bar{\rho}$, the agent picks an experiment $u$ that maximizes
$$\mathscr{P}_i(\vct{\rho},u) := {\sum_{j\neq i}\tilde{\rho}_jD(p_i^u||p_j^u)},$$
where $\tilde{\rho}_j = {\rho_j}/{1-\rho_i}.$ This strategy can be interpreted as the first player's best-response when the second player uses the mixed strategy $\tilde{{\rho}}_j$ to select an alternate hypothesis. Note that the mixed strategy $\vct{\alpha}_i^*$ used in OPE is an equilibrium strategy for the maximizing player.

\subsection{Simulation Setup}
To simulate these heuristics, we first consider a simple setup with three hypotheses and two queries. The conditional distributions $p_i^u(y)$ for each of these queries are illustrated in Figure \ref{table1}.

\begin{figure}[h]
%\vspace{-0.19in}
\centering
 \subfloat[][Query $u^1$]{
 \begin{tabular}{ | l | c | c |}
 \hline
   & $y = 0$& $y= 1$ \\
  \hline
  $h_0$ & 0.8 & 0.2 \\
  $h_1$ & 0.2 & 0.8 \\
  $h_2$ & 0.8 & 0.2\\
  \hline
\end{tabular}
 }
 \subfloat[][Query $u^2$]{
 \begin{tabular}{ | l | c | c |}
 \hline
   & $y = 0$& $y= 1$ \\
  \hline
  $h_0$ & 0.8 & 0.2 \\
  $h_1$ & 0.8 & 0.2 \\
  $h_2$ & 0.2 & 0.8\\
  \hline
\end{tabular}
 }
 \caption{Conditional distributions $p_i^u(y)$ for each query}
 \label{table1}
\end{figure}
The queries are designed such that when $\rv{H} = h_0$, the agent is forced to make both queries $u^1$ and $u^2$. This is because hypotheses $h_0$ and $h_2$ are indistinguishable under query $u^1$ and similarly, hypotheses $h_0$ and $h_1$ are indistinguishable under query $u^2$. However, when $\rv{H} = h_1$, the agent can eliminate $h_0$ and $h_2$ simultaneously using query $u^1$ alone and similarly, when  $\rv{H} = h_2$, the agent only needs to perform the query $u_2$.

\begin{figure}[]

%\centering
\hspace{-0.16in}\includegraphics[width=1.14\columnwidth]{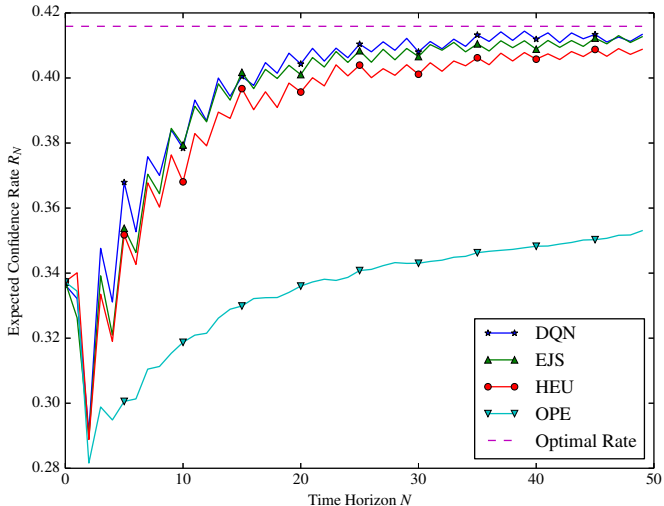}%

\caption{Evolution of expected confidence rate $R_N$ under hypothesis $h_0$ in the first setup with queries $u^1$ and $u^2$. Note the subpar performance of OPE in this setup.}
\label{plot1}
\end{figure}

We observe that all the four heuristics manage to learn this scheme of query selection. However, the rate at which confidence is maximized is different for each heuristic. We illustrate the evolution of expected confidence rate $R_N$ under hypothesis $h_0$ in Figure \ref{plot1}. The heuristics DQN, EJS and HEU come very close to the maximum achievable rate. OPE eventually achieves maximal rate but very slowly.

In the second experimental setup, we include two additional queries $u^3$ and $u^4$ characterized by the distributions in Figure \ref{table2}. When $\rv{H} = h_0$ the queries $u^3$ and $u^4$ together can eliminate at a much faster rate than $u^1$ and $u^2$. Intuitively, this is because when the agent performs $u^3$ and observes $y=1$, the belief on $h_1$ decreases drastically because $y=1$ is extremely unlikely under hypothesis $h_1$. Similarly, $u^4$ is very effective in eliminating $h_2$.

\begin{figure}[h!]
%\vspace{-0.19in}
\centering
 \subfloat[][Query $u^3$]{
 \begin{tabular}{ | l | c | c |}
 \hline
   & $y = 0$& $y= 1$ \\
  \hline
  $h_0$ & 0.8 & 0.2 \\
  $h_1$ & $1-\delta$ & $\delta$ \\
  $h_2$ & 0.8 & 0.2\\
  \hline
\end{tabular}
 }
 \subfloat[][Query $u^4$]{
 \begin{tabular}{ | l | c | c |}
 \hline
   & $y = 0$& $y= 1$ \\
  \hline
  $h_0$ & 0.8 & 0.2 \\
  $h_1$ & 0.8 & 0.2 \\
  $h_2$ & $1-\delta$ & $\delta$\\
  \hline
\end{tabular}
 }
 \caption{Conditional distributions $p_i^u(y)$ for each query. Here, $\delta = 0.0000001$.}
 \label{table2}
\end{figure}

\begin{figure}[]

%\centering
\hspace{-0.17in}\includegraphics[width=1.14\columnwidth]{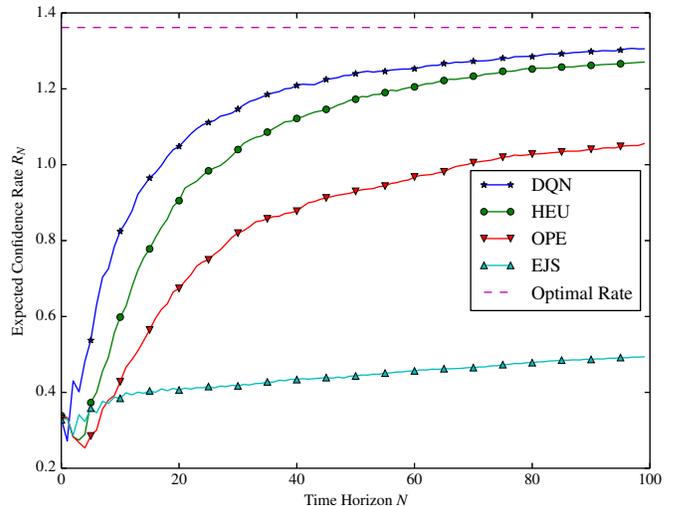}%

\caption{Evolution of expected confidence rate $R_N$ under hypothesis $h_0$ in the second setup with additional queries $u^3$ and $u^4$. Note the subpar performance of OPE and EJS in this setup.}
\label{plot2}
\end{figure}

The evolution of expected confidence rate under hypothesis $h_0$ with additional experiments $u^3$ and $u^4$ is shown in Figure \ref{plot2}. The heuristics DQN, HEU and OPE select queries $u^3$ and $u^4$ under hypothesis $h_0$. But the greedy heuristic EJS usually selects only $u^1$ and $u^2$ and fails to realize that queries $u^3$ and $u^4$ are more effective under hypothesis $h_0$. The greedy EJS approach fails because queries $u^3$ and $u^4$ are constructed in such way that they are optimal over longer horizons but are sub-optimal over shorter horizons. Thus the assumption required for asymptotic optimality of EJS in \cite{naghshvar2012extrinsic} does not hold in this setup. This also demonstrates that DQN is not simply selecting its queries greedily and manages to learn the long-term consequences of selecting queries.

\section{Conclusion}\label{conc}
In this paper, we considered the problem of active sequential hypothesis testing. We defined a notion of confidence, called Bayesian log-likelihood ratio and reformulated the hypothesis testing problem as a confidence maximization problem which can be seen as an infinite-horizon, average-reward MDP over a finite-dimensional belief space. We proposed a deep reinforcement learning based policy design framework for this MDP. We also proposed a heuristic based on a KL-divergence zero-sum game. Using numerical experiments, we compared these heuristics with those in prior works and demonstrated that our designed heuristics perform significantly better than existing methods in some scenarios.
%\begin{figure}[]
%
%%\centering
%\hspace{-0.14in}\includegraphics[width=1.1\columnwidth]{greedypic.eps}%
%
%\caption{An illustration of the regions $\mathcal{R}_1$ and $\mathcal{R}_2$}
%\label{greedyfig}
%\end{figure}

\section*{Acknowledgment}

This research was supported, in part, by National Science Foundation under Grant NSF CNS-1213128, CCF-1410009, CPS-1446901, Grant ONR N00014-15-1-2550, and Grant AFOSR FA9550-12-1-0215. We also thank Ashutosh Nayyar for his contribution in formulating and solving the MDP.

%\ifCLASSOPTIONcaptionsoff
%  \newpage
%\fi



%\bibliographystyle{IEEEtran}
\bibliographystyle{IEEEbib}
\bibliography{strings,refs}

% or


%

\appendices
%
%\section{Proof of Lemma \ref{levelset}}\label{sec:levelsetlemma}
%\section{Q-learning Algorithm for Finite State and Action Spaces}

\section{Recurrent Neural Network Architecture}\label{rnn}
\begin{figure}[]

%\centering
\includegraphics[width=1\columnwidth]{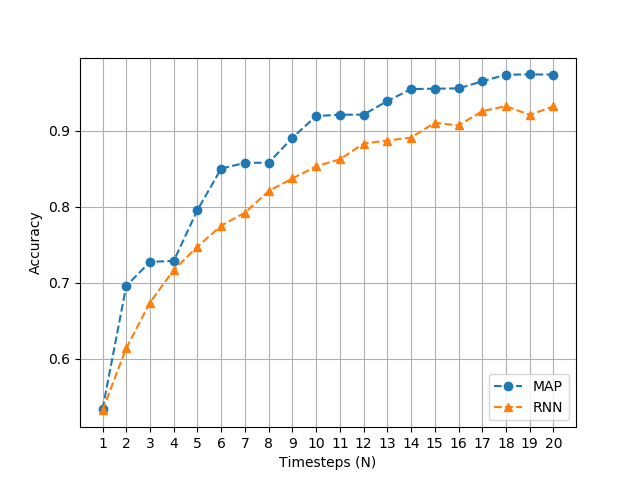}%

\caption{This plot compares the performance of the RNN network vs the MAP rule.}
\label{mapplot}
\end{figure}

\begin{figure}[]

\centering
\includegraphics[width=\columnwidth]{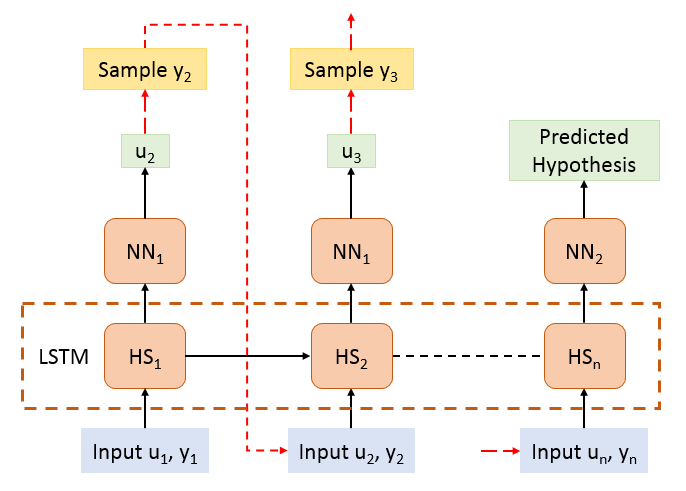}%

\caption{The LSTM network with query selection.}
\label{ripoff}
\end{figure}

The first goal is to verify if the internal state of an LSTM can maintain hypothesis information. The model is a simpler version of the recurrent network shown in Figure \ref{ripoff}, which takes a sequence of random queries and its results as input. This model is compared against Maximum A Posteriori (MAP) rule for hypothesis classification which is optimal for any input, in Figure \ref{mapplot}. The performance of LSTM comes close to that of MAP, which clearly shows that its hidden state maintains hypothesis information. 

We examine if LSTMs can learn query selection as well. Our model architecture in Figure \ref{ripoff} predicts a query at each time-step which in turn is used to produce an input for the next time-step. True hypothesis is provided for training the model, but optimal query selection for any time-step is unknown. The model is expected to learn this implicitly. There are two practical issues with this architecture. Query result is produced by a non-differentiable black-box making the output and input of consecutive time-steps disconnected. This prevents explicit learning of query selection. However, it is known that recurrent networks can learn implicit tasks \cite{sabir2017implicit}. Second, query selection is a soft decision made by the model, whereas a discrete decision is preferred. Experiments show that the model fails to learn query selection.

An improvement to this architecture can be made if hard decisions can be incorporated in a model. A discrete decision from the model also solves the problem of explicit query selection, since the output and input at consecutive time-steps can be connected. This direction of research leads to reinforcement learning, which requires further investigation.



%\end{IEEEbiography}
%% insert where needed to balance the two columns on the last page with
%%\newpage


%\vfill


\end{document}